\documentclass[aps,prl,twocolumn,amsmath,amssymb]{revtex4}

\usepackage{graphicx}
\usepackage{dcolumn}
\usepackage{bm}
\usepackage{epsfig}
\usepackage{epsfig}
\usepackage{amsmath}
\usepackage{amsfonts}

\newcommand{\ba}{\begin{eqnarray}}
\newcommand{\ea}{\end{eqnarray}}
\newcommand{\be}{\begin{equation}}
\newcommand{\ee}{\end{equation}}
\newcommand{\bea}{\begin{eqnarray}}
\newcommand{\eea}{\end{eqnarray}}

\usepackage{theorem}

\theorembodyfont{\rmfamily}

\theoremstyle{break}

\def\QED{~\rule[-1pt]{5pt}{5pt}\par\medskip}

\begin{document}


\title{Universal time scaling for Hamiltonian parameter estimation}
\author{Haidong Yuan}
\email{hdyuan@mae.cuhk.edu.hk}
\affiliation{Department of Mechanical and Automation Engineering, The Chinese University of Hong Kong, Shatin, Hong Kong}

\author{Chi-Hang Fred Fung}
\email{chffung.app@gmail.com}
\affiliation{
Canada Research Centre, Huawei Technologies Canada, Ontario, Canada,\\
Huawei Noah's Ark Lab, Hong Kong Science Park, Shatin, Hong Kong,
}

\date{\today}

\begin{abstract}
Time is a valuable resource and it seems intuitive that longer time should lead to better precision in Hamiltonian parameter estimation. However recent studies have put this intuition into question, showing longer time may even lead to worse estimation in certain cases. Here we show that the intuition can be restored if coherent feedback controls are included. By deriving asymptotically optimal feedback controls we present a quantification of the maximal improvement feedback controls can provide in Hamiltonian parameter estimation and show a universal time scaling for the precision limit of Hamiltonian parameter estimation under the optimal feedback scheme.

\end{abstract}

\maketitle

The implementation of quantum technology usually requires a full and precise information about the parameters of system evolution, which makes quantum Hamiltonian parameter estimation a crucial problem. An important task of Hamiltonian parameter estimation is to find out the ultimate achievable precision limit with given resources and design schemes that attain it. Typically Hamiltonian parameter estimation is achieved by preparing some initial quantum state $\rho_0$ and letting it evolve under the Hamiltonian $H(x)$, through the evolution $\rho_x=U_x\rho_0U^\dagger_x$, where $U_x=e^{-iH(x)T}$, the unknown parameter in the Hamiltonian is imprinted on $\rho_x$, one then can estimate the parameter by measuring $\rho_x$. This problem is well studied in quantum metrology when the Hamiltonian is in the multiplication form of the parameter $H(x)=xH$, it is known that in this case the optimal strategy is to prepare the initial state as $\frac{|\lambda_{max}\rangle +|\lambda_{min}\rangle}{\sqrt{2}},$
 where $|\lambda_{max(min)}\rangle$ is the eigenvector of $H$ for the maximum(minimum) eigenvalue, the standard deviation of the optimal unbiased estimator of $x$ then scales as $\frac{1}{\sqrt{nJ}}$, here $n$ is the number that the process is repeated and $J=(\lambda_{max}-\lambda_{min})^2T^2$ is the maximal quantum Fisher information, where $\lambda_{max(min)}$ is the maximum(minimum) eigenvalue of $H$ and $T$ is the time that the Hamiltonian acts on initial states\cite{Giovannetti2011}. In this case the standard deviation of the estimation scales as $\frac{1}{T}$ which shows that longer time always leads to better precision, consistent with our intuition. However for general Hamiltonian $H(x)$, recent studies have shown different time scalings\cite{pang2014}. For example, if a Hamiltonian takes the form $H(x)=B[\cos(x)\sigma_1+\sin(x)\sigma_3]$, where $\sigma_1=\left(\begin{array}{cc}
      0 & 1  \\
      1 & 0 \\
          \end{array}\right)$, $\sigma_2=\left(\begin{array}{cc}
      0 & -i  \\
      i & 0 \\
          \end{array}\right)$ and $\sigma_3=\left(\begin{array}{cc}
      -1 & 0  \\
      0 & 1 \\
          \end{array}\right)$ are Pauli matrices, the maximum quantum Fisher information in this case is then $J=4\sin^2 BT$, which oscillates with time\cite{pang2014,Liu2014}. Thus longer time may even lead to worse estimation for general Hamiltonian, this contradicts with our intuition.

In this article, we will show that the intuition can be restored when we include feedback controls. 
By presenting an asymptotically optimal feedback scheme for general Hamiltonian parameter estimation, we give a quantification of the maximal improvement feedback controls can provide in Hamiltonian parameter estimation. We show that under the optimal scheme the precision limit displays a universal time scaling $\frac{1}{T}$ which is independent of the form of the Hamiltonian. In this article we focus on single parameter estimation, generalization to multiple parameters is possible but is beyond the scope of this article.
 The methods developed previously in \cite{pang2014,Liu2014} to compute maximal quantum Fisher information for general Hamiltonians are quite invovled and hard to incorporate feedback controls. Here we use a tool developed in our recent work which is computationally efficient and convenient to include feedback controls, which we recapture here briefly\cite{Yuan2015}. The precision of estimating $x$ from quantum states $\rho_x$ is related to the Bures distance between $\rho_x$ and its neighboring states $\rho_{x+dx}$\cite{HELS67,HOLE82,BRAU94,BRAU96},
 \begin{equation}
\label{eq:BJ}
d^2_{Bures}(\rho_x,\rho_{x+dx})=\frac{1}{4}J(\rho_x)dx^2,
\end{equation}
here the Bures distance $d_{Bures}$ is defined as
\begin{equation}
\label{eq:Bures}
d_{Bures}(\rho_1,\rho_2)=\sqrt{2-2F_B(\rho_1,\rho_2)},
 \end{equation}
where $F_B(\rho_1,\rho_2)=\sqrt{\rho_1^{\frac{1}{2}}\rho_2\rho_1^{\frac{1}{2}}}$ is the fidelity. Thus maximizing the quantum Fisher information $J(\rho_x)$ is equivalent to maximizing the square of Bures distance between $\rho_x$ and its neighboring states.

If the evolution is governed by $U_x=e^{-iH(x)T}$ with a general Hamiltonian $H(x)$, then $\rho_x=U_x\rho_0U_x^\dagger$ and $\rho_{x+dx}=U_{x+dx}\rho_0U_{x+dx}^\dagger$, thus
\begin{eqnarray}
\aligned
&\max_{\rho_0} d^2_{Bures}(\rho_x,\rho_{x+dx})\\
&=\max_{\rho_0} d^2_{Bures}(U_x\rho_0U_x^\dagger,U_{x+dx}\rho_0U_{x+dx}^\dagger)\\
&=\max_{\rho_0} [2-2F_B(U_x\rho_0U_x^\dagger,U_{x+dx}\rho_0U_{x+dx}^\dagger)]\\
&=\max_{\rho_0}[2-2F_B(\rho_0, U_x^\dagger U_{x+dx}\rho_0 U_{x+dx}^\dagger U_x)]\\
&=2-2\min_{\rho_0}F_B(\rho_0,U'\rho_0 U'^\dagger),
\endaligned
\end{eqnarray}
where  $U'=U_x^\dagger U_{x+dx}$. Let $e^{-i\theta^{U'}_j}$ be eigenvalues of $U'$, where $\theta^{U'}_j\in(-\pi,\pi]$ for $1\leq j\leq d$(here $d$ denotes the dimension of $U'$), which we will call eigen-angles of $U'$, and arrange $\theta^{U'}_{\max}=\theta^{U'}_1\geq \theta^{U'}_2\geq \cdots \geq \theta^{U'}_d=\theta^{U'}_{\min}$ in decreasing order. Then
$\min_{\rho_0}F_B(\rho_0,U'\rho_0 U'^\dagger)=\cos\frac{\theta^{U'}_{\max}-\theta^{U'}_{\min}}{2}$ if $\theta^{U'}_{\max}-\theta^{U'}_{\min}\leq \pi$\cite{Fung2}. We denote $C_{TE}(U)=\frac{\theta^{U}_{\max}-\theta^{U}_{\min}}{2}$ for a given unitary operator,
then from Eq.(\ref{eq:BJ}) we get
\begin{eqnarray}
\label{eq:maxQFIphi1}
\aligned
\max_{\rho_0} J&=\lim_{dx\rightarrow 0}\frac{8(1-\cos\frac{\theta^{U'}_{\max}-\theta^{U'}_{\min}}{2})}{dx^2}\\
&=\lim_{dx\rightarrow 0}\frac{8[1-\cos C_{TE}(U_x^\dagger U_{x+dx})]}{dx^2}.
\endaligned
\end{eqnarray}
The precision limit is then given by
\begin{eqnarray}
\aligned
\label{eq:Precision1}
\delta \hat{x}&\geq\frac{1}{\sqrt{n\max_{\rho_0}J[U_x\rho_0U^\dagger_x]}}\\
&=\frac{1}{\lim_{dx\rightarrow 0} \frac{\sqrt{8[1-\cos C_{TE}(U_x^\dagger U_{x+dx})]}}{\mid dx\mid}\sqrt{n}},
\endaligned
\end{eqnarray}
where $n$ is the number that the measurement procedures are repeated.
If $U_x$ is continuous with respect to $x$, then when $dx\rightarrow 0$, $U'=U^\dagger(x)U(x+dx)\rightarrow I$, thus $\theta^{U'}_{\max},\theta^{U'}_{\min}\rightarrow 0$ and $\frac{\theta^{U'}_{\max}-\theta^{U'}_{\min}}{2}\rightarrow 0$. We then get
\begin{eqnarray}
\label{eq:maxQFIphi}
\aligned
\max_{\rho_0} J&=\lim_{dx\rightarrow 0}\frac{8(1-\cos\frac{\theta^{U'}_{\max}-\theta^{U'}_{\min}}{2})}{dx^2}\\
&=\lim_{dx\rightarrow 0}\frac{16\sin^2\frac{\theta^{U'}_{\max}-\theta^{U'}_{\min}}{4}}{dx^2}\\
&=\lim_{dx\rightarrow 0}\frac{(\theta^{U'}_{\max}-\theta^{U'}_{\min})^2}{dx^2}\\
&=\lim_{dx\rightarrow 0}\frac{4C_{TE}^2(U_x^\dagger U_{x+dx})}{dx^2},
\endaligned
\end{eqnarray}
which gives the ultimate precision limit
\begin{eqnarray}
\aligned
\label{eq:Precision}
\delta \hat{x}&\geq\frac{1}{\sqrt{n\max_{\rho_0}J}}\\
&=\frac{1}{\lim_{dx\rightarrow 0} \frac{2C_{TE}(U_x^\dagger U_{x+dx})}{\mid dx\mid}\sqrt{n}}.
\endaligned
\end{eqnarray}

For Hamiltonians in the multiplication form $H(x)=xH$, $U_x=e^{-ixHT}$, $U_x^\dagger U_{x+dx}=e^{-iHTdx}$, in this case $$C_{TE}(U_x^\dagger U_{x+dx})=(\lambda_{\max}-\lambda_{\min})T\mid dx\mid ,$$  Eq.(\ref{eq:Precision}) then recovers the well-known formula\cite{Giovannetti2011}
\begin{eqnarray}
\aligned
\delta \hat{x}&\geq\frac{1}{\sqrt{n}(\lambda_{\max}-\lambda_{\min})T}.
\endaligned
\end{eqnarray}

For general Hamiltonians, this also provides a straight forward way of calculating maximum quantum Fisher information. We will demonstrate it through an example, which will also be used later to show the gain of feedback controls. Consider the Hamiltonian $H(x)=B[\cos(x)\sigma_1+\sin(x)\sigma_3]$, where $x$ is the interested parameter, it can represent the direction of a magnetic field\cite{pang2014,Liu2014}. The Hamiltonian can be written compactly as $H(x)=B[\vec{a}(x)\cdot \vec{\sigma}]$, where $a_1(x)=\cos(x),a_2(x)=0, a_3(x)=\sin(x)$.
 If it evolves with time $T$, then $U_x=e^{-iH(x)T}=e^{-iBT[\vec{a}(x)\cdot \vec{\sigma}]}$, in this case
\begin{eqnarray}
\aligned
U'&=U^\dagger_xU_{x+dx}\\
&=e^{iBT[\vec{a}(x)\cdot \vec{\sigma}]}e^{-iBT[\vec{a}(x+dx)\cdot \vec{\sigma}]}.\\
\endaligned
\end{eqnarray}
With a simple calculation one can get 
\begin{eqnarray}
\aligned
e^{iBT[\vec{a}(x)\cdot \vec{\sigma}]}e^{-iBT[\vec{a}(x+dx)\cdot \vec{\sigma}]}=e^{iB'[\vec{a'}\cdot \vec{\sigma}]},\\
\endaligned
\end{eqnarray}
here $\vec{a'}$ is a unit vector and 
\begin{eqnarray}
\aligned
\cos B'
=&\cos^2(BT)+\cos(dx)\sin^2(BT)\\
=&\cos^2(BT)+(1-\frac{dx^2}{2})\sin^2(BT)+O(dx^3)\\
=&1-\sin^2(BT)\frac{dx^2}{2}+O(dx^3).
\endaligned
\end{eqnarray}
Since the eigenvalues of $e^{iB'(\vec{a'}\cdot\vec{\sigma})}$ are $e^{\pm iB'}$, we have $\theta^{U'}_{\max}=B'$ and $\theta^{U'}_{\min}=-B'$, thus $\frac{\theta^{U'}_{\max}-\theta^{U'}_{\min}}{2}=B'$. From Eq.(\ref{eq:maxQFIphi1}) we then get $$\max J=\lim_{dx\rightarrow 0} 8\frac{1-\cos B'}{dx^2}=4\sin^2(BT).$$ This is consistent with previous studies\cite{pang2014,Liu2014},  however our method makes the computation much simpler.


Next we are going to include feedback controls and show that the intuition that time is a valuable resource can be restored. We first prove a useful property of $C_{TE}(U)$: given two n-dimensional unitary operators $U_1$ and $U_2$, if $C_{TE}(U_1)+C_{TE}(U_2)\leq \pi$, then $C_{TE}(U_1U_2)\leq C_{TE}(U_1)+C_{TE}(U_2)$. To see this, let $U_1=e^{-iH_1}$ and $U_2=e^{-iH_2}$ where the eigenvalues of $H_1$ and $H_2$ are in $[-\pi,\pi]$. From Thompson's theorem\cite{Thom}, there exists two unitary operators $V_1$ and $V_2$ such that
\begin{eqnarray}
U_1U_2=e^{-iH_1}e^{-iH_2}=e^{-i(V_1H_1V_1^\dagger+V_2H_2V_2^\dagger)}.
\end{eqnarray}
Denote $H_3=V_1H_1V_1^\dagger+V_2H_2V_2^\dagger$, we have $\lambda(H_3)\prec \lambda(H_1)+\lambda(H_2)$\cite{Fan}, here $\lambda(H)$ denotes the vector whose entries are the eigenvalues of $H$, arranged in decreasing order, i.e., $\lambda_1(H)\geq \lambda_2(H)\geq \cdots \geq \lambda_n(H)$ here $\lambda_i(H)$ denotes an eigenvalue of $H$ at the i-th entry of $\lambda(H)$, '$\prec$' denotes the majorization relation\cite{Bhatia}, $\lambda(H_3)\prec \lambda(H_1)+\lambda(H_2)$ means
\begin{eqnarray}
\sum_{i=1}^k\lambda_i(H_3)\leq \sum_{i=1}^k \lambda_i(H_1)+\sum_{i=1}^k\lambda_i(H_2)
\end{eqnarray}
for $1\leq k \leq n-1$ and $$\sum_{i=1}^n\lambda_i(H_3)=\sum_{i=1}^n \lambda_i(H_1)+\sum_{i=1}^n\lambda_i(H_2).$$
From this it is easy to see that
\begin{eqnarray}
\lambda_1(H_3)\leq \lambda_1(H_1)+\lambda_2(H_2),\\
\lambda_n(H_3)\geq \lambda_1(H_n)+\lambda_2(H_n).
\end{eqnarray}
Thus $\lambda_1(H_3)-\lambda_n(H_3)\leq \lambda_1(H_1)-\lambda_n(H_1)+\lambda_1(H_2)-\lambda_n(H_2)$. Since $C_{TE}(e^{-iH})=\frac{\lambda_1(H)-\lambda_n(H)}{2}$, this shows
\begin{equation}
\label{eq:CTE}
C_{TE}(U_1U_2)\leq C_{TE}(U_1)+C_{TE}(U_2).
\end{equation}

\begin{figure}
\begin{minipage}{.5\textwidth}
  \centering
  \includegraphics[width=.9\linewidth]{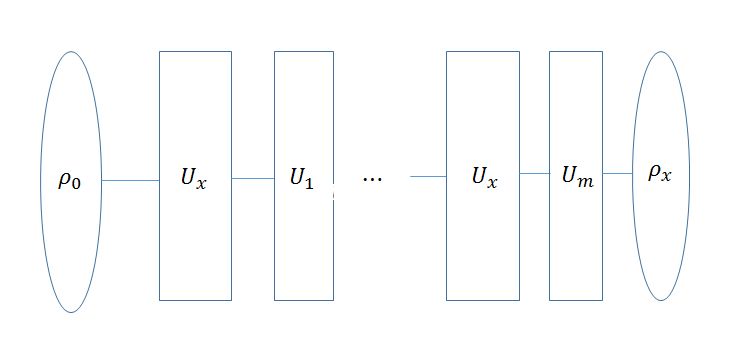}
  \caption{Hamiltonian parameter estimation with feedback controls.}
  \label{figfeedback}
\end{minipage}
\label{fig:test}
\end{figure}
We are now ready to derive optimal feedback controls. As shown in Fig.\ref{figfeedback} under the feedback scheme the evolution is interspersed with feedback coherent controls.
The total evolution can be described by $$U_{mt}(x)=U_t(x)U_1U_t(x)U_2\cdots U_t(x)U_m,$$ here $U_t(x)=e^{-iH(x)t}$ with $t=\frac{T}{m}$, $U_1, U_2,\cdots ,U_m$ are coherent controls. Here we assume the controls are evenly interspersed, this is mostly for simplicity of notations, the analysis can be generalized to uneven case straightforwardly. 
We will first derive optimal controls for the case of $m=2$, same strategy works in the general case. When $m=2$, $U_{2t}(x)=U_t(x)U_1U_t(x)U_2$, then
\begin{eqnarray}
\label{eq:align}
\aligned
&U_{2t}^\dagger(x) U_{2t}(x+dx)\\
=&U_2^\dagger U^\dagger_t(x)U_1^\dagger U^\dagger_t(x) U_t(x+dx)U_1U_t(x+dx)U_2\\
=&U_2^\dagger U^\dagger_t(x)U_1^\dagger [U^\dagger_t(x) U_t(x+dx)]U_1[U_t(x)U^\dagger_t(x)]\\
&U_t(x+dx)U_2\\
=&U_2^\dagger (U^\dagger_t(x)U_1^\dagger) [U^\dagger_t(x) U_t(x+dx)](U_1U_t(x))\\
&[U^\dagger_t(x)U_t(x+dx)]U_2.\\
\endaligned
\end{eqnarray}
From Eq.(\ref{eq:maxQFIphi}) we know finding the maximal quantum Fisher information is equivalent as finding the maximum value of $C_{TE}[U_{2t}^\dagger(x) U_{2t}(x+dx)]$. Since $C_{TE}[U_{2t}^\dagger(x) U_{2t}(x+dx)]$ is determined by the eigenvalues of $U_{2t}^\dagger(x) U_{2t}(x+dx)$ and $U_2$ does not change the eigenvalues in this case, so it can be chosen as any unitary. This is reasonable as measurements on the output states have already implicitly assumed additional controls. We divide the rest of operators into two parts, $(U^\dagger_t(x)U_1^\dagger) [U^\dagger_t(x) U_t(x+dx)](U_1U_t(x))$ and $U^\dagger_t(x) U_t(x+dx)$, apply the inequality (\ref{eq:CTE}) we get
\begin{eqnarray}
\aligned
C_{TE}&[U_{2t}^\dagger(x) U_{2t}(x+dx)]\leq C_{TE}[U^\dagger_t(x) U_t(x+dx)]\\
&+C_{TE}[(U^\dagger_t(x)U_1^\dagger) [U^\dagger_t(x) U_t(x+dx)](U_1U_t(x))]\\
&=2C_{TE}[U^\dagger_t(x) U_t(x+dx)],
\endaligned
\end{eqnarray}
where the last equality we used the fact that $(U^\dagger_t(x)U_1^\dagger) [U^\dagger_t(x) U_t(x+dx)](U_1U_t(x))$ has the same eigen-angles as $U^\dagger_t(x) U_t(x+dx)$.
One obvious choice of control that saturates the equality is $U_1=U^\dagger_t(x)$, as it aligns the eigenvalues of the two parts and the corresponding maximal and minimal eigen-angles add up.  In this case $U_{2t}^\dagger(x) U_{2t}(x+dx)=U_2^\dagger[U^\dagger_t(x) U_t(x+dx)]^2U_2$, $C_{TE}[U_{2t}^\dagger(x) U_{2t}(x+dx)]=2C_{TE}[U_t^\dagger(x) U_t(x+dx)].$

Note that this is not in contradiction with previous result which showed that feedback controls do not help when $H(x)=xH$\cite{GIOV06}. The reason feedback controls do not help in that case is $U(x)$ commutes with $U(x+dx)$, thus $(U^\dagger(x)U_1^\dagger) [U^\dagger(x) U(x+dx)](U_1U(x))$ is already equal to $U^\dagger(x) U(x+dx)$ without any controls(i.e., when $U_1=I$), so the eigen-angles are already aligned. However for general Hamiltonians, $U(x)$ and $U(x+dx)$ in general do not commute, in that case feedback controls are useful to increase the precision limit.

\begin{figure}
  \centering
  \includegraphics[width=\columnwidth]{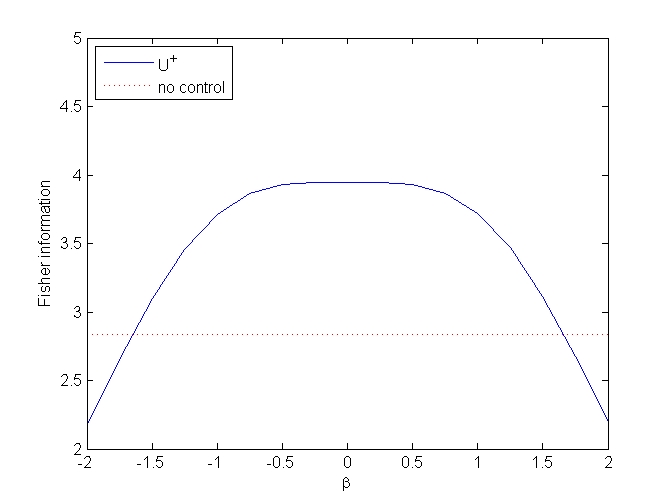}
  \caption{Quantum Fisher information for parameter estimation of the Hamiltonian $H(x)=B[\cos(x)\sigma_1+\sin(x)\sigma_3]$ with $B=1$,$T=1$ and $t=1/5$, i.e., the evolution is interspersed with 5 controls. The controls are taken as $U^\dagger_t(\hat{x})$ with $\hat{x}=(1+\beta)x$ and the real value of $x$ is assumed to be $1$. In this case the feedback controls gain as long as $\beta$ is in $(-1.66, 1.66)$.}
  \label{fig:beta}
\end{figure}

This analysis can be extended to general $m$ straightforwardly with
\begin{equation}
\label{eq:TEcontrol}
C_{TE}[U_{mt}^\dagger(x)U_{mt}(x+dx)]\leq mC_{TE}[U^\dagger_t(x)U_t(x+dx)],
\end{equation}
where the equality can be saturated with the controls $U_1=U_2=\cdots =U_{m-1}=U^\dagger_t(x)$ and arbitrary $U_m$.
In practice the true value of $x$ is not known beforehand, thus an estimated value $\hat{x}$ has to be used for the feedback controls, and the controls $U_1=U_2=\cdots =U_{m-1}=U^\dagger_t(\hat{x})$ need to be updated adaptively. The maximum quantum Fisher information is achievable asymptotically when $\hat{x}\rightarrow x$ \cite{FUJI06,NAGA88,NAGA89,HAYA08}. In the non-asymptotical regime, with a reasonable estimate such strategy can also gain.
For example with $H(x)=B[\cos(x)\sigma_1+\sin(x)\sigma_3]$ and consider the feedback scheme with a total evolution time $T=mt$ and let the controls $U_1=U_2=\cdots =U^\dagger_t(\hat{x})$ with $\hat{x}=(1+\beta)x$, here $\beta$ represents the error of the estimate. In Fig.(\ref{fig:beta}) we plotted the Fisher information with different $\beta$, it can be seen that such feedback controls gain for a broad range of $\beta$. When $\beta=0$, the controls leads to the maximum quantum Fisher information which equals to $4m^2\sin^2\frac{BT}{m}$, this is asymptotically achievable and much higher than the maximum Fisher information without feedback controls. When $m$ is sufficiently large, $\sin \frac{BT}{m}\doteq\frac{BT}{m}$, the maximum quantum Fisher information reaches $ 4B^2T^2$, which means that the ultimate precision limit scales as $\frac{1}{T}$.

Such time scaling is actually universal that holds for any $H(x)$ under the optimal feedback scheme, which we will now show. 
Assume the evolution is interspersed with $m$ controls, $$U_{mt}(x)=U_t(x)U_1U_t(x)U_2\cdots U_t(x)U_m,$$ here $U_t(x)=e^{-iH(x)t}$ with $t=\frac{T}{m}$. Under the optimal strategy $U_1=U_2=\cdots =U_{m-1}=U^\dagger_t(x)$, $C_{TE}[U_{mt}^\dagger(x)U_{mt}(x+dx)]$ attains its maximal value $mC_{TE}[U^\dagger_t(x) U_t(x+dx)]$. If $m$ is taken sufficiently large, then $t=\frac{T}{m}$ can be sufficiently small, in this case
\begin{eqnarray}
\aligned
U^\dagger_t(x) U_t(x+dx)&=e^{iH(x)t}e^{-iH(x+dx)t}\\
&\doteq e^{-i\frac{[H(x+dx)-H(x)]T}{m}},
\endaligned
\end{eqnarray}
thus
$$C_{TE}[U^\dagger_t(x) U_t(x+dx)]\doteq \frac{[\lambda_{\max}(x,dx)-\lambda_{\min}(x,dx)]T}{m},$$ here $\lambda_{\max(\min)}(x,dx)$ denotes the maximum(minimum) eigenvalue of $H(x+dx)-H(x)$. Thus when $m$ is sufficiently large $$mC_{TE}[U^\dagger_t(x) U_t(x+dx)]\doteq[\lambda_{\max}(x,dx)-\lambda_{\min}(x,dx)]T,$$ which gives the maximal quantum Fisher information
\begin{eqnarray}
\label{eq:T2}
\aligned
\max J(\rho_x)
=& 4T^2\frac{[\lambda_{\max}(x,dx)-\lambda_{\min}(x,dx)]^2}{dx^2},
\endaligned
\end{eqnarray}
 the ultimate precision limit is then given by $$\delta \hat{x}\geq \frac{1}{\sqrt{nJ}}= \frac{1}{2\sqrt{n}\lim_{dx\rightarrow 0}\frac{\lambda_{\max}(x,dx)-\lambda_{\min}(x,dx)}{\mid dx\mid}T},$$ this shows the time scaling for the ultimate precision limit is universal. 

\begin{figure}
  \centering
  \includegraphics[width=\columnwidth]{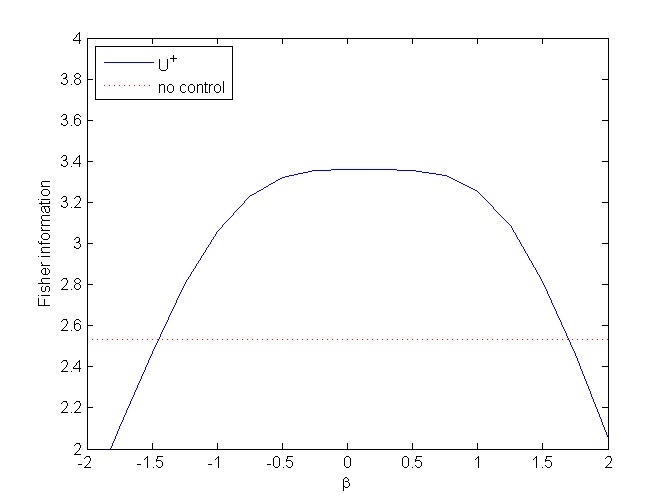}
  \caption{Quantum Fisher information for parameter estimation of
 noisy channel having Kraus operators
$K_1(x)=\sqrt{\frac{1+\eta}{2}}U(x)$  and $K_2(x)=\sqrt{\frac{1-\eta}{2}}\sigma_3 U(x)$
where
$U(x)=\exp(-i H(x)t)$,
$\eta=0.8^{1/5}$, and
  the Hamiltonian $H(x)=B[\cos(x)\sigma_1+\sin(x)\sigma_3]$ with $B=1$, $T=1$ and $t=1/5$, i.e., the evolution is interspersed with 5 controls. The controls are taken as $U^\dagger_t(\hat{x})$ with $\hat{x}=(1+\beta)x$ and the real value of $x$ is assumed to be $1$. In this case the feedback controls gain as long as $\beta$ is in $(-1.46, 1.70)$.}
  \label{fig:beta-noisy}
\end{figure}




\noindent { \emph{Summary and outlook: }}we derived an asymptotically optimal feedback scheme for Hamiltonian parameter estimation and showed that under this scheme the ultimate precision limit has a universal time scaling, this restored the intuition that time is always a valuable resource when the evolution is unitary. At the presence of noise, such feedback strategy is still useful in increasing the precision limit, for example at the presence of dephasing noises as shown in Fig.\ref{fig:beta-noisy}, the feedback controls still gain for a broad range of $\beta$. Recently it has been shown that for a noisy evolution with the Hamiltonian $H(x)=xH$, quantum error correcting techniques can be used to extend the time scaling to $\frac{1}{T}$ if the noise is correctable\cite{Arrad2014,Kessler2014,Ozeri2013,Dur2014}. It is straightforward to combine those techniques with the optimal feedback controls here to show that universal time scaling also holds for arbitrary Hamiltonian under correctable noises. The combined method is expected to lead to better precision in general, however the optimality will generally be lost with a simple concatenation of these two techniques(quantum error correcting is not optimal in general), finding out the optimal feedback controls at the presence of noises will be a future research direction.


\begin{thebibliography}{99}
\bibitem{Giovannetti2011}
Giovannetti, V., Lloyd, S. \& Maccone, L.
{\em Nature Photonics.} {\bf 5}, 222 (2011).
\bibitem{GIOV06}
Giovannetti, V., Lloyd, S. \& Maccone, L.,
Quantum metrology.
{\em Phys. Rev. Lett.} {\bf 96}, 010401 (2006).
\bibitem{pang2014}
Pang, S.S., \& Brun, T.A.,
Quantum metrology for a general Hamiltonian parameter.
{\em Phys. Rev. A.} {\bf 90}, 022117 (2014)

\bibitem{Liu2014}
Liu, J., Jing, X. \& Wang, X.G.,
Quantum metrology with unitary parametrization processes,
Scientific Reports, 5, 8565 (2015).
\bibitem{Yuan2015}
Yuan, H.D. \& Fung, C.-H. F.,
manuscript.
\bibitem{HELS67}
Helstrom, C. W.,
Quantum Detection and Estimation Theory.
(Academic Press, New York, 1976).
\bibitem{HOLE82}
Holevo, A. S.,
Probabilistic and Statistical Aspect of Quantum Theory.
(North-Holland, Amsterdam, 1982).

\bibitem{BRAU94}
Braunstein, S. L. \& Caves, C.  M.,
Statistical distance and the geometry of quantum states.
{\em Phys. Rev. Lett.} {\bf 72}, 3439 (1994).

\bibitem{BRAU96}
Braunstein, S. L., Caves, M. C. \& Milburn, G. J.,
Generalized Uncertainty Relations: Theory, Examples,  and Lorentz Invariance.
{\em Annals of Physics} {\bf 247}, 135-173 (1996).


\bibitem{Fisher}
Fisher, R. A.
 {\em Proc. Cambr. Phil. Soc.}
{\bf 22}, 700 (1925).

\bibitem{CRAM46}
Cram\'{e}r, H.,
Mathematical Methods of Statistics.
(Princeton University, Princeton NJ, 1946).
\bibitem{Rao}
Rao, C. R.
{\em Bull. Calcutta Math. Soc.} {\bf 37}, 81 (1945).

\bibitem{Chau2011}
Chau, H. F.,
{\em Quant. Inf. Compu.} {\bf 11}, 0721 (2011).
\bibitem{Fung1}
Fung, C.-H. F. \& Chau, H. F.,
{\em Phys. Rev. A} {\bf 88}, 012307 (2013).
\bibitem{Fung2}
Fung, C.-H. F. \& Chau, H. F.,
{\em Phys. Rev. A} {\bf 90}, 022333 (2014).
\bibitem{FUJI06}
Fujiwara, A.,
Strong consistency and asymptotic efficiency for adaptive quantum estimation problems.
{\em J. Phys. A: Math. Gen.} {\bf 39}, 12489-12504 (2006).

\bibitem{NAGA88}
Nagaoka, H.,
An asymptotic efficient estimator for a one-dimensional parametric model of quantum statistical operators.
{\em Proc. Inf. Symp. on Inform. Theory}, 198 (1998).
\bibitem{Berry2000}
Berry, D. W. \& Wiseman, H. M.,
Optimal states and almost optimal adaptive measurements for quantum interferometry.
{\em Phys. Rev. Lett.} {\bf 85}, 5098 (2000).
\bibitem{NAGA89}
Nagaoka, H.,
On the parameter estimation problem for quantum statistical models.
{\em Proc. 12th Symp. on Inform. Theory and its Appl.}, 577-82 (1989).

\bibitem{HAYA08}
Hayashi, M. \& Matsumoto, K.,
Asymptotic performance of optimal state estimation in qubit system.
{\em J.  Math. Phys.} {\bf 49}, 102101 (2008).

\bibitem{Thom}
R. C. Thompson, Linear and Multilinear Algebra {\bf 19}, 187
(1986).
\bibitem{Fan}
K. Fan, Proc. Nat. Acam. Sci. {\bf 35}, 131 (1949).
\bibitem{Bhatia} R.Bhatia,
{"Matrix Analysis"}
  Springer-Verlag, {New York},
  1997.
  
\bibitem{Dur2014} D\"ur, W., Skotiniotis, M., Fr\"owis, F. \&  Kraus B.
Improved quantum metrology using quantum error correction.
\textit{Phys. Rev. Lett.} \textbf{112}, 080801 (2014).

\bibitem{Arrad2014} Arrad, G., Vinkler, Y., Aharonov, D. \& Retzker, A.
Increasing Sensing Resolution with Error Correction.
\textit{Phys. Rev. Lett.} \textbf{112}, 150801 (2014).

\bibitem{Kessler2014} Kessler, E. M., Lovchinsky, I., Sushkov, A. O. \& Lukin, M. D.
Quantum error correction for metrology.
\textit{Phys. Rev. Lett.} \textbf{112}, 150802 (2014).

\bibitem{Ozeri2013} Ozeri, R.
Heisenberg limited metrology using quantum error-correction codes.
arXiv:1310.3432.

%
\end{thebibliography}
\end{document}